\pdfoutput=1

\documentclass[11pt]{article}

\usepackage[preprint]{acl}

\usepackage{times}
\usepackage{latexsym}

\usepackage[T1]{fontenc}

\usepackage[utf8]{inputenc}

\usepackage{microtype}

\usepackage{inconsolata}

\usepackage{graphicx}
\usepackage{amsmath}
\usepackage{amssymb}
\usepackage{booktabs}
\usepackage{multirow}
\usepackage{colortbl}
\usepackage{ascmac}
\usepackage{makecell} 

\title{Bias-Adjusted LLM Agents for Human-Like Decision-Making via Behavioral Economics}

\author{Ayato Kitadai\textsuperscript{*}\\
  School of Engineering\\
  The University of Tokyo\\ \And
  Yusuke Fukasawa\\
  School of Engineering\\
  The University of Tokyo\\
  \textsuperscript{*}\textbf{Correspondence:} \href{mailto:a.kitadai@css.t.u-toyo.ac.jp}{a.kitadai@css.t.u-toyo.ac.jp} \\ \And
  Nariaki Nishino\\
  School of Engineering\\
  The University of Tokyo\\
}

\begin{document}
\maketitle
\begin{abstract}
Large language models (LLMs) are increasingly used to simulate human decision-making, but their intrinsic biases often diverge from real human behavior—limiting their ability to reflect population-level diversity.
We address this challenge with a persona-based approach that leverages individual-level behavioral data from behavioral economics to adjust model biases.
Applying this method to the ultimatum game—a standard but difficult benchmark for LLMs—we observe improved alignment between simulated and empirical behavior, particularly on the responder side.
While further refinement of trait representations is needed, our results demonstrate the promise of persona-conditioned LLMs for simulating human-like decision patterns at scale.
\end{abstract}

\section{Introduction}
The world envisioned by \citet{Simon-1959}, where insights from computer science accelerate understanding of human decision-making, is becoming a reality with the rise of Large Language Models (LLMs). 
While agent-based modeling attracted attention to this idea~\citep{Gilbert-2000}, it struggled to incorporate social, cultural, and cognitive complexities~\citep{Sims-1980}, limiting its adoption in fields like microeconomics.
In contrast, LLM-driven agents (LLM agents) differ from traditional rule-based systems, fueling expectations for more flexible and realistic simulations~\citep{Grossman-2023, Gao-2024}.
Indeed, LLM agents have already succeeded in replicating macro-level human behavior~\citep{Park-2023, Li-2024-acl, qian-etal-2024-chatdev}, raising hopes for their application in individual decision-making.

\begin{figure}
    \centering
    \includegraphics[width=\linewidth]{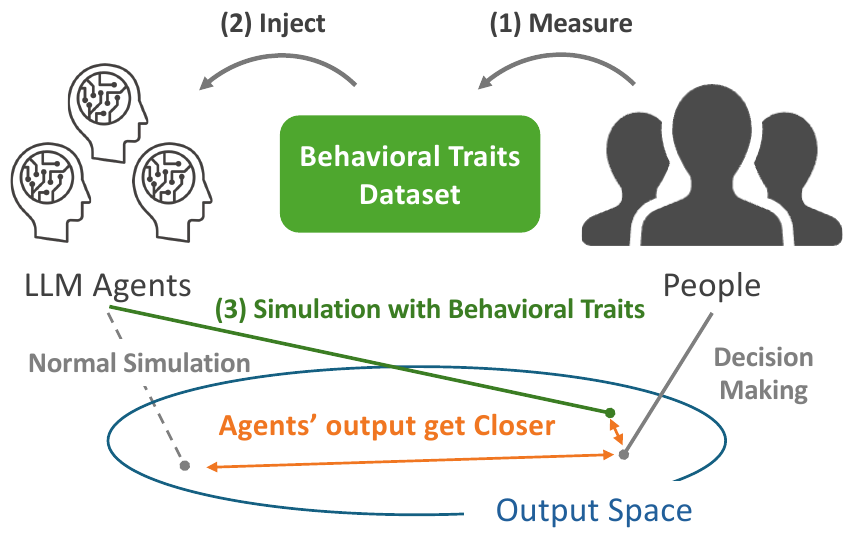}
    \caption{Overview of our method. Behavioral traits from the Econographics dataset are injected into LLMs as personas, enabling decision-making that more closely aligns with human behavior.}
    \label{fig:concept}
\end{figure}

However, existing research has shown that the decision-making of LLM agents often deviates from that of humans. 
A large body of work has compared human responses and those of LLM agents, particularly in the context of questionnaire-based studies (e.g. \citealp{Argyle-2023}), psychological experiments (e.g. \citealp{Hagendorff-2023-NCS}), and economic experiments (e.g. \citealp{Aher-2023}). 
While some studies have reported consistency between the two (e.g. \citealp{Li-2024, Filippas-2024}), the majority have identified similarities while also pointing out notable discrepancies (e.g. \citealp{Dominguez-2024, Mei-2024, Xie-2024, tjuatja-etal-2024-llms}). 

A key challenge in using LLMs for human behavior modeling lies in their inherent biases, which originate from the training data and learning processes~\citep{blodgett-etal-2020-language, Lin-2024}.
These models exhibit distinct cognitive, psychological, and cultural biases~\citep{taubenfeld-etal-2024-systematic, liu-etal-2024-empirical, fang-etal-2024-born, kumar-etal-2024-subtle, naous-etal-2024-beer, bang-etal-2024-measuring, an-etal-2024-large}, yet aligning them with human behavior remains a major obstacle~\citep{kang-etal-2024-large}.
Such biases raise concerns about using LLM agents as substitutes for human participants, as they risk failing to reflect diverse human identities~\citep{Grossman-2023, Harding-2024}.
Although efforts to create human-like LLM agents have gained attention in NLP (e.g., \citealp{chuang-etal-2024-simulating, chuang-etal-2024-beyond}), adjusting these biases to more accurately reflect human behavior remains an open problem.

To address this, we leverage behavioral economics to adjust biases in LLM agents and better align their decisions with human behavior.
Behavioral economics refines traditional economic theory by quantifying human biases and bridging the gap between normative models and actual behavior.
Similarly, its insights may help close the gap between LLM agents and real human populations.
Modifying persona attributes—such as cultural background, personality traits, and behavioral biases—can influence LLM outputs~\citep{jiang-etal-2024-personallm, Phelps-2025}.
To enable more human-like agents, we use Econographics~\citep{Data-econographics}, a dataset of behavioral indicators that allows persona settings to reflect real-world bias distributions.
We evaluate this approach by testing whether it can reproduce results from one-shot ultimatum games, which LLMs have struggled to replicate~\citep{Kitadai-2023, Mei-2024}.

This study makes two main contributions. 
First, we propose a novel method for adjusting LLM biases using individual-level behavioral traits to better align their decisions with human behavior.
Second, we demonstrate that this approach enables LLMs to replicate human responses in one-shot ultimatum games—a task previously difficult for these models~\citep{Kitadai-2023, Mei-2024}.
These results offer a foundation for simulating human populations in behavioral research.

\section{Methodology}
\subsection{Ultimatum Game}
\begin{figure}[htbp]
    \centering
    \includegraphics[width=\linewidth]{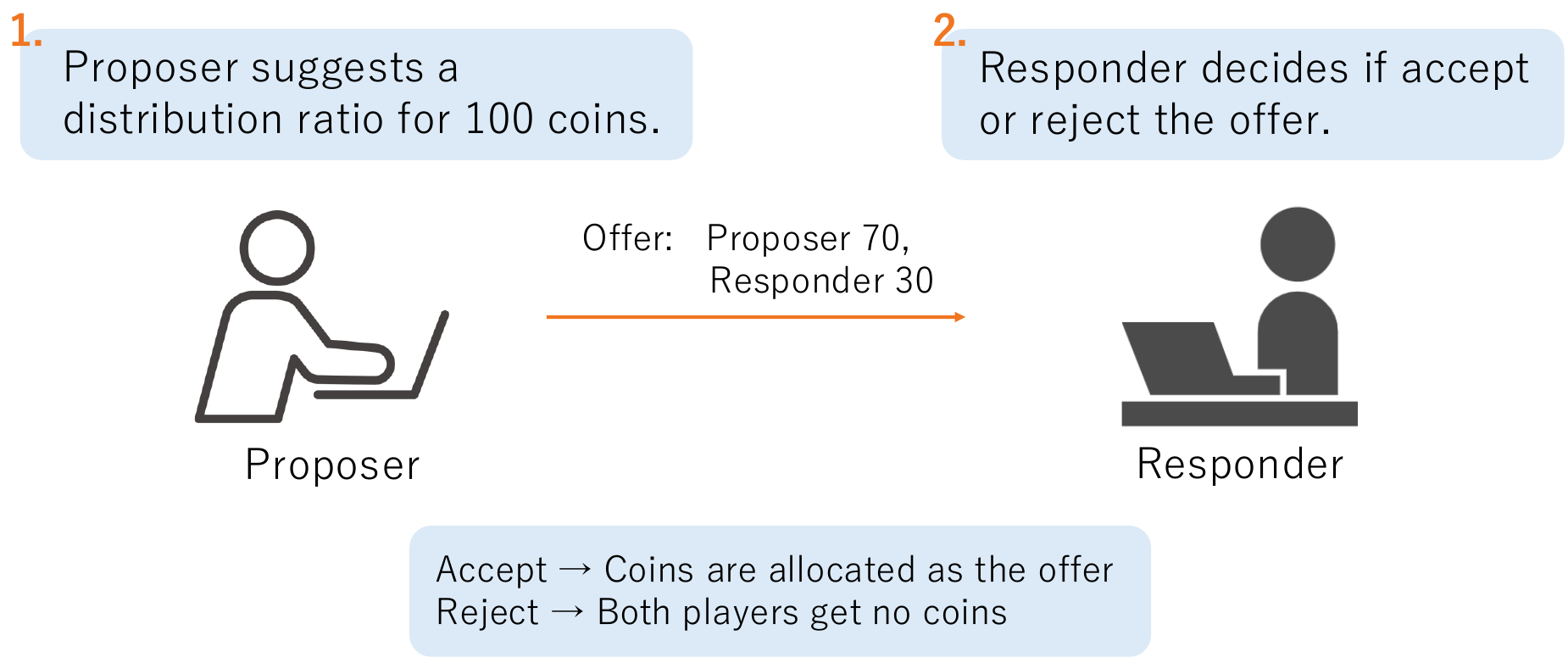}
    \caption{Intuitive description of ultimatum game}
    \label{fig:ultimatum}
\end{figure}

The ultimatum game~\citep{Guth-1982} involves two players dividing $100$ coins (Figure~\ref{fig:ultimatum}).
The proposer suggests an allocation to the responder ($s_1 \in \{0,\dots,100\}$), who then chooses to accept or reject ($s_2 \in \{\text{accept}, \text{reject}\}$).
If accepted, the proposed split is implemented; if rejected, both receive nothing:
\begin{equation*}
    (u_1, u_2) = 
    \begin{cases} 
        (100 - s_1, s_1) &\text{if } s_2 = \text{accept},\\
        (0, 0) &\text{if } s_2 = \text{reject} .
    \end{cases} 
\end{equation*}

Classical game theory predicts that responders should accept any nonzero offer, prompting proposers to offer as little as possible.
Empirically, however, humans tend to offer close to $50$ coins, and responders often reject low offers—highlighting fairness concerns and punishment behavior.
We adopt the data from \citet{Lin-2020} as a benchmark, as it exhibits these widely observed behavioral patterns.

\subsection{Persona Settings with Econographics}
To characterize LLM agents based on behavioral economic traits, we utilized \textit{Econographics} from \citet{Chapman-2023}. They aimed to establish an empirical foundation for a more comprehensive understanding of decision theory by analyzing correlation patterns among various behavioral indicators.
Specifically, they conducted a monetary-incentivized survey involving $1000$ participants of diverse ages in the United States and extracted $21$ behavioral indicators, which were then reduced to six principal components through principal component analysis (PCA) as shown in Table~\ref{tab:econographics}.

\begin{table*}
\centering
\renewcommand{\arraystretch}{1.2} 
\resizebox{\textwidth}{!}{
\begin{tabular}{ccccccc}
\hline
\textbf{Components}
& \textbf{Generosity}
& \begin{tabular}{@{}c@{}} \textbf{Punishment}\\\textbf{(Impulsivity)} \end{tabular}
& \begin{tabular}{@{}c@{}} \textbf{Inequality}\\\textbf{Aversion / WTP} \end{tabular}
& \textbf{WTA}
& \textbf{Uncertainty}
& \textbf{Overconfidence} \\
\hline
\textbf{$21$ Indicators}
& \begin{tabular}{@{}c@{}}
\cellcolor{red!30}Reciprocity: High\\
Reciprocity: Low\\
Altruism\\
Trust
\end{tabular}
& \begin{tabular}{@{}c@{}}
\cellcolor{red!30}Anti-social Punishment\\
Pro-social Punishment\\
Patience
\end{tabular}
& \begin{tabular}{@{}c@{}}
Dislike Having More\\
Dislike Having Less\\
WTP\\
\cellcolor{red!30}Risk Aversion: CR (Certain)\\
Risk Aversion: CR (Lottery)
\end{tabular}
& \begin{tabular}{@{}c@{}}
WTA\\
\cellcolor{red!30}Risk Aversion: Gains\\
Risk Aversion: Losses\\
Risk Aversion: Gain/Loss
\end{tabular}
& \begin{tabular}{@{}c@{}}
\cellcolor{red!30}Ambiguity Aversion\\
Compound Lottery Aversion
\end{tabular}
& \begin{tabular}{@{}c@{}}
Overestimation\\
Overplacement\\
\cellcolor{red!30}Overprecision
\end{tabular} \\
\hline
\end{tabular}
} 
\caption{Summary of principal components identified by \citet{Chapman-2023}, based on PCA of $21$ behavioral indicators. Each column corresponds to one component, and rows list associated indicators. Red-highlighted items are most strongly loaded on each component and used in our $6$ trait configuration.}
\label{tab:econographics}
\end{table*}

A key advantage of using microdata is that each LLM agent can be assigned the exact behavioral and demographic attributes of a real individual from the dataset. This enables us to construct a heterogeneous agent population that more accurately reflects the diversity observed in human decision-making. 
In our study, each of the $1000$ LLM agents is assigned a unique persona based on the individual-level measured values of all $21$ behavioral indicators, as well as Cognitive Reflection Test (CRT) scores, age, gender, and country of residence from the dataset~\citep{Data-econographics}.

\subsection{Simulation Procedure}
To ensure consistency with human experiments, we used the exact instructions from \citet{Lin-2020}, where participants divided $100$ coins with three LLMs.
Following \citet{Kitadai-2023}, agents were told each coin was worth $0.10$ dollars, but were not explicitly instructed to maximize profit.


To systematically evaluate the impact of behavioral traits on agent decision-making, we implemented three persona-setting configurations: no persona settings (baseline), all $21$ behavioral indicators from the dataset, and just $6$ key behavioral indicators, representing each of the principal components identified by PCA in \citet{Chapman-2023}.
The prompting procedure consisted of four steps: (1) persona assignment, (2) game explanation, (3) decision query, and (4) output format specification.
We generated $1000$ agents per condition and fixed the temperature to zero for all outputs.
For fair comparison, responder agents evaluated the same $1000$ offers sampled from \citet{Lin-2020}.

\section{Results and Discussion}
\begin{figure*}[htbp]
    \centering
    \includegraphics[width=\linewidth]{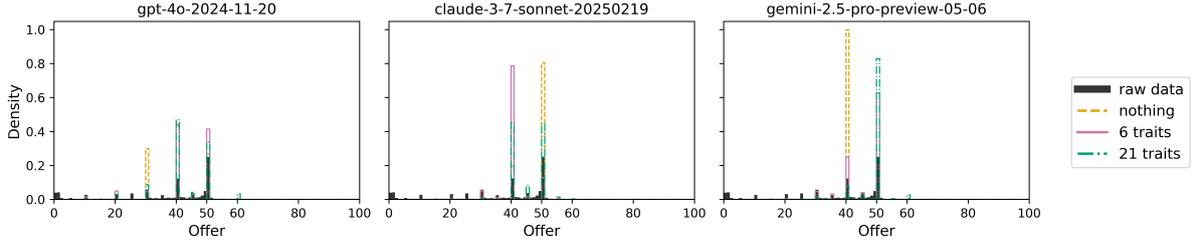} 
    \caption{Distribution of offers proposed by LLM agents under different persona configurations. Each panel shows results for a different LLM. Black bars represent human experimental data from \citet{Lin-2020}, while colored curves correspond to simulations under no persona (orange, dashed), $6$ trait (blue), and $21$ trait (green) conditions.}
    \label{fig:result_proposer} 
\end{figure*}

\begin{figure*}[htbp]
    \centering
    \includegraphics[width=\linewidth]{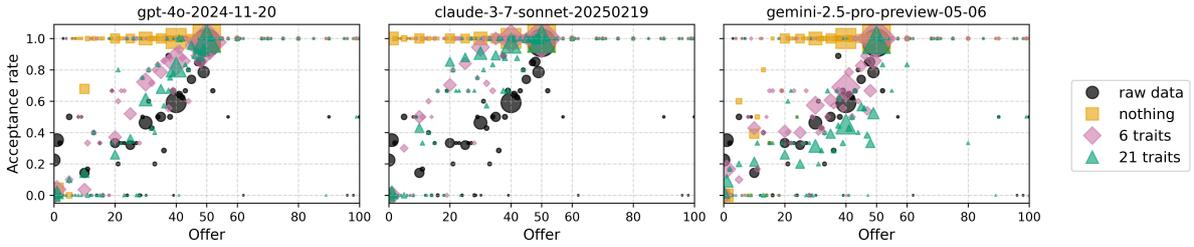} 
    \caption{Acceptance rates of LLM agents in the responder role, plotted against offer size. Each panel shows results for a different LLM. Black points represent human data from \citet{Lin-2020}; colored bubbles show LLM simulations under no persona (orange, dashed), $6$ trait (blue), and $21$ trait (green) settings. Bubble size indicates the frequency of each offer in the dataset.}
    \label{fig:result_responder} 
\end{figure*}

Figure~\ref{fig:result_proposer} shows frequency of each value proposed to the opponent, while figure~\ref{fig:result_responder} visualizes responder acceptance rates to each proposed coins as bubble plots where the bubble size reflects offer frequency.
Panels in each figure corresponds to different LLMs.
Black plots represent human experimental results from \citet{Lin-2020}, while colored plots depict LLM-based simulations.

\paragraph{Proposer Side}
Without persona settings, proposer behavior diverges across models, revealing distinct internal biases.
GPT-4o centers around $40$ coins, Claude peaks at $50$, and Gemini offers almost exclusively $40$—indicating model-specific fairness priors.
Persona conditioning modifies these tendencies.
GPT-4o shifts its mode toward $50$, suggesting increased fairness; Claude spreads offers more broadly below $50$, resembling human variability; and Gemini, previously rigid, centers sharply at $50$ under both trait settings.

\paragraph{Responder Side}
Without persona input, all models accept nearly any offer above $20$ coins, resulting in flat acceptance curves that deviate from human patterns, which show gradual increases by offer size.
Introducing behavioural traits lowers acceptance rates for unfair offers and yields more human-like, monotonic curves—especially between $0$ and $50$ coins.
Overall, personas help move LLM responders away from purely rational strategies toward fairness-sensitive behavior that better reflects human judgement.

\paragraph{Quantitative Evaluation}
We assess alignment with human data by computing Wasserstein distances between simulation outputs and empirical distributions (Table~\ref{tab:quant_comp}).
Across models, persona conditioning reduces distance.
Proposer-side gains are modest, reflecting partial alignment, while responder-side improvements are more substantial, particularly for GPT-4o (from $0.289$ to $0.090$).
Notably, GPT-4o and Claude benefit most from the full $21$ traits, whereas Gemini achieves better alignment with the six-trait setting—suggesting greater sensitivity to input dimensionality.

\begin{table}[htbp]
    \centering
    \resizebox{1.0\columnwidth}{!}{
    \begin{tabular}{c c c c}
        \toprule
        \textbf{Model} & \textbf{Persona} & \textbf{Proposer} & \textbf{Responder} \\
        \midrule

        \multirow{3}{*}{\makecell{\small{gpt-4o}\\\small{-2024-11-20}}}
        & Nothing        & $1.13$ & $0.289$ \\
        & $6$ Traits    & $0.961$ & $0.137$ \\
        & $21$ Traits    & \textbf{0.923} & \textbf{0.090} \\
        \midrule

        \multirow{3}{*}{\makecell{\small{claude-3-7}\\\small{-sonnet-20250219}}}
        & Nothing        & $1.25$ & $0.332$ \\
        & $6$ Traits    & \textbf{1.08} & $0.213$ \\
        & $21$ Traits    & $1.09$ & \textbf{0.141} \\
        \midrule

        \multirow{3}{*}{\makecell{\small{gemini-2.5-pro}\\\small{-preview-05-06}}}
        & Nothing        & $1.49$ & $0.227$ \\
        & $6$ Traits    & \textbf{1.00} & \textbf{0.082} \\
        & $21$ Traits    & $1.15$ & $0.172$ \\
        \bottomrule
    \end{tabular}
    }
    \caption{Wasserstein distances between simulated and human behavioral distributions in the ultimatum game. Each row shows results for one LLM under different persona conditions. Values are scaled by $10^{-2}$; lower values indicate closer alignment with human data. Bolded values indicate the lowest distance for each model.}
    \label{tab:quant_comp}
\end{table}

\paragraph{Model-Specific Behavioral Effects}
The differential impact of persona conditioning across roles reflects the cognitive demands of each task.
Responders evaluate fixed offers, tied to traits like inequality aversion and reciprocity.
These are captured in the Econographics dataset, making even low-dimensional traits effective in aligning LLM behaviour with human-like fairness sensitivity.

Proposer behavior, in contrast, requires anticipating a responder’s reaction—a generative task involving strategic reasoning and social expectations.
This decision-making is likely harder to steer using static trait information alone. 
While some models, such as GPT-4o, showed partial alignment with human data under trait conditioning, overall improvements were limited and varied across models.
These mixed results suggest that capturing proposer behavior may require richer or more dynamic persona inputs beyond the current trait set.

Model-specific differences also emerged.
GPT-4o and Claude appear to benefit from high-dimensional trait vectors, integrating them in a way that stabilizes behavior.
Gemini, on the other hand, performs better with reduced inputs.
One possible explanation is that Gemini more directly incorporates trait information into its reasoning process.
When given too many traits, it may reflect them too literally—resulting in exaggerated behavioral responses, such as overly low acceptance rates.
In contrast, GPT-4o and Claude may implicitly moderate or interpolate across conflicting signals.

These findings underscore the importance of tailoring persona design to the characteristics of each LLM.
Trait conditioning is a promising tool for human-aligned simulation, but its effectiveness depends not only on trait selection, but also on how each model interprets and integrates such inputs.

\section{Conclusion}
In this paper, we proposed a persona-based characterization of LLM agents using a diverse set of behavioral indicators from behavioral economics.
Leveraging these persona settings improved the replication of one-shot ultimatum game results, particularly on the responder side.

The remaining discrepancy in proposer behavior suggests that the $21$ indicators used may not fully capture human-like traits.
Concepts in other fields such as the Big Five personality traits are known to complementarily influence human behavior~\citep{Becker-2012}. 
Expanding the dataset with more diverse indicators could enable richer persona settings, further aligning LLM agent responses with human behavior.

While this study focuses on single-shot decisions, our broader goal is to build toward fully interactive, agentic systems capable of participating in dynamic economic environments.
Such systems could eventually enable policy simulations, virtual field experiments, and population-scale modeling grounded in real human behavior.
Although further research is needed, our findings highlight the potential of LLM agents to simulate human-like behavior at scale.

\section*{Limitations}

This study has several limitations. First, particularly regarding the transparency of the LLMs used. Our simulations rely exclusively on some of commercial models, without extending validation to a broader range of open-source LLMs. 

Additionally, there are limitations in both internal and external validity that must be considered.  
In terms of internal validity, a key limitation is the discrepancy between the participant populations in the datasets used. The behavioral indicator data in \citet{Chapman-2023} and economic experiment data in \citet{Lin-2020} were collected from different groups. To rigorously assess the effectiveness of the proposed approach, it would be ideal to collect both behavioral indicators and experimental outcomes from the same participant panel, ensuring a more controlled and comprehensive evaluation.  

Regarding external validity, this study focuses solely on the one-shot ultimatum game, which constrains the generalizability of our findings. Economic research encompasses a variety of classical experimental paradigms, such as the prisoner’s dilemma, public goods games, and auction experiments. Moreover, real-world decision-making extends beyond these well-established frameworks, involving newly proposed institutional designs, more complex experimental setups, and dynamic decision-making contexts. Future research should explore a wider range of economic games and decision-making scenarios to evaluate the robustness of LLM-based simulations more comprehensively.

Moreover, in this study, we did not compare our proposed method with alternative approaches such as fine-tuning. As a result, it remains unclear how our method performs relative to existing techniques or whether it offers complementary advantages. This is a limitation of the current work and should be addressed in future research.

\bibliography{anthology,custom}

\appendix
\section{Prompts Construction}
It has long been known that in economic experiments and questionnaire surveys involving actual human subjects, the wording, context, and choice of words in questions and instructions can significantly affect the outcomes (e.g. \citealp{ Bertrand-2001}). 
\citet{Camerer-2016} emphasized that the availability of instructions used during experiments is crucial for replicating economic experiments accurately.

In addtion, \citet{Kitadai-2023} attempted to reproduce the result of the ultimatum game by LLM agents focusing on the instruction regarding monetary incentives.
As a result, they found that not giving explicit instructions to maximize profits, specifying the amount to be received by the participants at the end of the experiment made the simulation result closer to the experimental result.

Based on these, we constructed the prompts mainly using the instruction actually utilized in the experiments in \citet{Lin-2020} with informing agents that each coin was worth $0.1$, without explicitly directing them to maximize payoffs. 
An example of prompt we used is shown in Table~\ref{tab:prompt_exp_prop_bias}.

\begin{table*}[htbp]
    \centering
    \begin{tabular}{p{3cm}p{10cm}}
    \toprule
        Structure & Prompt\\
    \midrule
        1. Agent Persona Setting & You embody a character with the following personality traits and demographics, each quantified by a value. A mean of $0$ and a standard deviation of $1$ are assumed for these values, where higher values signify a stronger inclination towards the corresponding trait:\\
        &\quad - Reciprocity:High : \textcolor{red}{$-0.1254847674509733$}\\
        &\quad - Anti-social Punishment : \textcolor{red}{$0.8201827598505876$}\\
        &\quad - Risk Aversion:Gains : \textcolor{red}{$0.1056063995538051$}\\
        &\quad - Risk Aversion: CR Certain : \textcolor{red}{$-0.6871043035159886$}\\
        &\quad - Ambiguity Aversion : \textcolor{red}{$-0.23977249139149004$}\\
        &\quad - Overprecision : \textcolor{red}{$0.48407670683235493$}\\
        &\quad - Age : \textcolor{red}{$50$}\\
        &\quad - Gender : \textcolor{red}{female}\\
        &\quad - CRT Score : \textcolor{red}{$1$} of 3\\
        &\quad - Country of Residence: US\\
        2. Explanation of the Game & Play the following game as someone of this personality. Here is the game description.\par A stack of coins is being divided between a proposer and a responder. The proposer decides how much to give the responder, and the responder decides whether or not to accept the offer. If the offer is accepted, the players split the money as the proposer suggested. If the offer is rejected, both parties receive no coin.\\
        3. Detailed Explanation of how each agent makes a decision & You are a proposer. $100$ coins will be divided. Each coin will be redeemed $0.1$ real-world dollar after the experiment. How would you suggest?\\
        4. Output Format Specification & Please tell me your decision in the following JSON format.\par
        $\{$\par
        \quad ``Reason'': ``Your explanation here'',\par
        \quad ``Responder'': ``Number of coins for the responder'',\par
        \quad ``Proposer'': ``Number of coins for yourself''\par
        $\}$\\
    \bottomrule
    \end{tabular}
    \caption{Prompt example for the proposer side with $6$ key behavioral indicators}\label{tab:prompt_exp_prop_bias}
\end{table*}

\end{document}